\documentclass[twocolumn, 10pt]{IEEEtran}
\usepackage{graphicx,color}  
\usepackage{sidecap} 
\usepackage{dcolumn}   
\usepackage{bm}        
\usepackage{amssymb}   
\usepackage{gensymb}    

\usepackage{ulem}
\usepackage{amsmath}
\usepackage{epstopdf}

\usepackage{siunitx} 

\newcommand{\um}{\,\si{\micro\meter}}

\newcommand{\C}{\,\si{\celsius}}

\newcommand{\uL}{\,\si{\micro\liter}}

\newcommand{\mM}{\,\si{\milli\textsc{m}}}

\newcommand{\ulh}{\si{\micro\liter\per\hour}}

\newcommand{\uM}{\si{\micro\textsc{m}}}
\newcommand{\nM}{\si{\nano\textsc{m}}}
\newcommand{\pM}{\si{\pico\textsc{m}}}

\usepackage[margin=2 cm]{geometry}

\usepackage{authblk}

\title{A DNA-encoded recipe to direct multi-stage colloidal assembly}

\author[1,2*]{Pepijn G. Moerman}
\author[1]{Chenghung Chou}
\author[3]{Thomas E. Videb\ae k}
\author[3]{W. Benjamin Rogers}
\author[1,$\dagger$]{Rebecca Schulman}
\affil[1]{Department of Chemical and Biomolecular Engineering, Johns Hopkins University, Baltimore, MD 21218, USA}
\affil[2]{Department of Chemical and Chemical Engineering, Eindhoven University of Technology, Eindhoven, 5612 AE, The Netherlands}
\affil[3]{Martin A. Fisher School of Physics, Brandeis University, Waltham, MA 02453, USA}
\affil[*]{Correspondence can be addressed to: p.g.moerman@tue.nl}
\affil[$\dagger$]{Correspondence can be addressed to: rschulm3@jhu.edu}
\date{}                     
\begin{document}
 \maketitle
 
\begin{abstract}
\textbf{In equilibrium self-assembly, microscopic building blocks spontaneously self-organize into stable structures as dictated by their interaction potentials, which limits the accessible structural features to those that correspond to global minima in free energy landscapes; they are often ordered and periodic on length scales comparable to the building block size. Coupling the assembly process to an exergonic reaction drives the system out of equilibrium so that an assembly pathway can be engineered to target a specific kinetically stabilized state, which in principle opens up a vast design space with access to diverse complex structures with features on multiple length scales. However, the question of how such features might be specifically targeted remains unanswered. Here, we explore this design space using a DNA-encoded recipe consisting of multiple biomolecular reactions that dictate the time-dependent binding strength and specificity of each type of subunit in the sample independently, which makes it possible to program an assembly pathway that leads to a kinetically trapped final state. With this kinetic control, we show that the same set of building blocks can form clusters with different final structures. These structures, with tunable core-shell compositions, have feature sizes much larger than the building block size and are governed by the DNA-encoded assembly kinetics. Contrasting global kinetic control strategies such as thermal annealing, tuning the timing of individual biomolecular reactions offers the opportunity to regulate how the activity of each separate co-assembling component of a large set varies over time, opening up the potential for morphogenesis-like assembly processes involving engineered species.}
\end{abstract}

\section{Introduction}
{T}he self-assembly of colloidal building blocks is a promising method for producing materials with structure on the nanometer to micrometer scales~\cite{Whitesides2002,Boles2016}, from which the materials can derive useful mechanical~\cite{Lossada2020}, optical~\cite{avci2018,cersonsky2021,He2020}, and transport properties~\cite{Doherty2009}. To attain a specific structure, the guiding principle is generally to sculpt the free energy landscape such that the target structure occupies the global minimum~\cite{Dijkstra2021}. For ordered structures, this entails designing building blocks whose interaction energies only compensate for the decrease in configurational entropy associated with the self-assembly process when the building blocks are in the target configuration~\cite{Torquato2009,Zhang2004}. The information that dictates the final structure is thus encoded in the building blocks' interactions, via their  specificity\cite{Hormoz2011,Wu2012,Zeravcic2014,Murugan2015} or directionality~\cite{Kraft2012,Chen2011,Wang2012}. Because this design principle provides precise control over the local structure of the particle assembly, it is well-suited for periodic materials, such as crystal lattices~\cite{Romano2012,Wang2015b,Laramy2019}. However, materials with structural or compositional variation on length scales larger than the interaction range are difficult to target through equilibrium assembly, and either require that each building block in the cluster is uniquely encoded~\cite{liu2016,Zeravcic2014} --- fully addressable assembly ---, that the lattice constant is much larger than at least one dimension of the building block size~\cite{hayakawa2024, wintersinger2023} or that there is an aggregate-size dependent binding energy, resulting from, for example, a building block curvature~\cite{Hayakawa2022} or an elastic defect energy~\cite{Hackney2023}. 

A bio-inspired alternative to targeting patterns on this mesoscopic length scale between the building block size and the system size is to couple the assembly processes to exergonic reactions~\cite{Ravensteijn2020}. This coupling both facilitates assembly into meta-stable structures that have higher free energies than the global minimum~\cite{Grzybowski2017}---the system's free energy may still decrease due to the exergonic reaction---and provides access to mesoscopic length scales that come from competing transport and reaction rates~\cite{Kondo2010}. For example, the length distribution of actin filaments~\cite{Pollard2000} and microtubules\cite{Hess2017} is controlled by an ATP hydrolyzation reaction that switches the protein building blocks between an assembly-active and inactive state. Contrary to external tools to control assembly pathways such as thermal annealing~\cite{Michele2013,Jo2024} or light patterning~\cite{Zhu2020}, the bio-inspired approach of molecular recipes can tune multiple processes with varied rates at once with independent levers.

DNA-coated colloidal building blocks make it possible to systematically explore this out-of-equilibrium design space because (1) the sequence selectivity and temperature-dependence of DNA hybridization enable tunable programmable interactions between DNA-coated particles~\cite{Mirkin1996, Alivisatos1996,Cui2022,Rogers2016,Geerts2010,Kahn2020, jacobs2025}, and (2) myriad commercially available enzymes exist that produce, degrade, or cleave oligonucleotides, enabling time-dependent interactions between colloidal particles by coupling the particles' surface properties to reaction networks~\cite{Grosso2022,Zadorin2017,Sharma2023}. 
Particles coupled to these enzymatic reactions can send and process nucleic acid signals, making it possible to control the lifetime of transient aggregates~\cite{Dehne2019,Sharma2023}, the frequency of oscillatory assembly~\cite{Dehne2021}, and the position of localized aggregation in a signal concentration gradient~\cite{Zadorin2017}. A way to explicitly dictate the time-dependence of interaction potentials of selective building blocks would make it possible to systematically explore assembly pathways and create stepwise rules to target specific structural features. 

Here, we present a strategy for controlling the colloidal assembly kinetics using a DNA-encoded recipe that sets the time-dependence of the binding strength and specificity of different types of colloids. The recipe consists of a set of template DNA strands that catalyze the growth of new single-stranded DNA domains on the DNA-coated particles~\cite{Moerman2023}. It is based on the Primer Exchange Reaction (PER)~\cite{kishi2018}, an isothermal, templated DNA polymerization reaction. We find that the templates, co-factors that do not assemble, control the rates of the polymerization reactions that in turn determine when each type of particle can bind to its binding partners. We show that with this control over the time-dependent interaction between the building blocks, we can program multiple assembly stages to produce kinetically trapped core-shell aggregates, where the structure of these aggegrates depends on the assembly pathway and not on the particles' properties after the assembly process; the same particles can assemble into qualitatively different structures when following different trajectories dictated by the template co-factors (Fig.~\ref{Fig1}a).  

\section*{Results}
\subsection*{Tunable assembly lag time}
\begin{figure}
\includegraphics[scale=1]{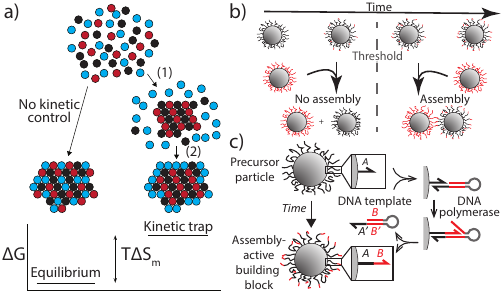}
\caption{\label{Fig1} a) Self-assembly in multiple stages can lead to kinetically trapped structures with compositional variation on length scales larger than the building block size. The core-shell structure has a higher free energy than the mixed equilibrium structure due to the entropy of mixing $\Delta S_m$. b) Slow polymerization of new DNA onto DNA-coated colloids results in delayed colloidal aggregation. The precursor particles are coated with DNA that is not complementary to DNA on any other particles in the system (black lines) and thus do not assemble. 9-nucleotide DNA domains that are complementary to the DNA on the co-assembler particles (red lines) can be  added to the precursor particles \textit{via} the primer exchange reaction (PER). The precursor particles become assembly-active particles and can bind to the co-assembler particles after sufficient new DNA domains (red) have been added to them. c) PER converts precursors to assembly-active particles by adding a specific ssDNA domain to the particle-grafted DNA using a hairpin-shaped catalytic template strand. The template's sequence controls which precursor particles get converted (sequence domain A) and what particles the assembly-active building block reacts with (sequence domain B); its concentration controls the reaction rate.}
\end{figure}
\begin{figure*}
\centering
\includegraphics[scale=1]{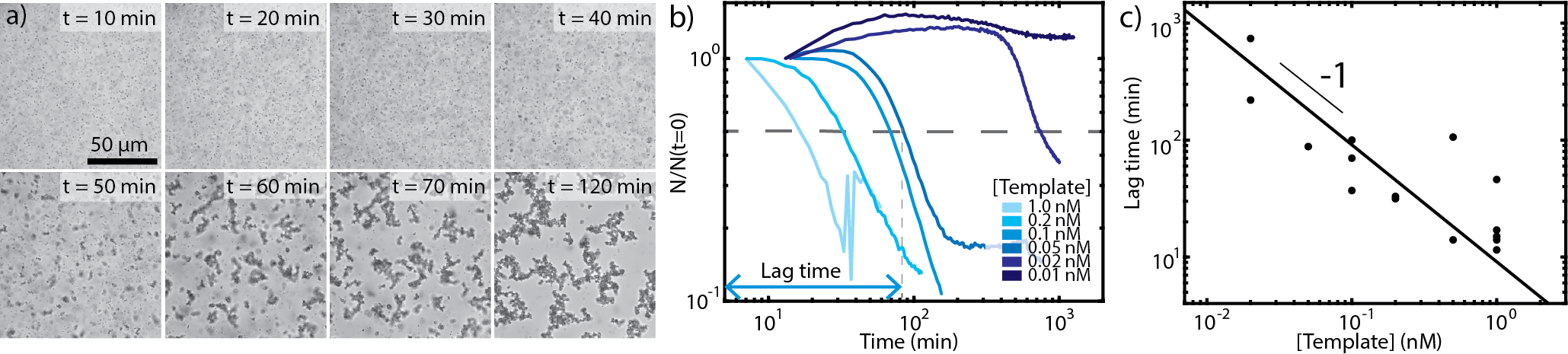}
\caption{\label{Fig2} a) Micrographs captured at different times during the delayed aggregation of DNA-coated colloids caused by a slow DNA polymerization reaction. The sample contains $10~\nM$ of template, which resulted in an assembly delay of approximately 50 minutes. b)  $N/N_{t=0}$, a measure of the fraction of unaggregated particles, as a function of time. The lag time decreases with increasing template concentration. The lag time is defined as the time it takes for $N/N(t=0)$ to drop below 0.5 (dashed line). The initial increase in N for the long-time experiments comes from slow sedimentation of the particles, which causes the number of particles in the field of view to increase. The fluctuations at the end are an imaging artifact from binarization of images with few large clusters where one pixel can cause the difference between one large and two smaller aggregates. c) The lag time as a function of template concentration. The lag time is approximately inversely proportional to the template concentration and depends on the typical reaction rate $\tau$ and the total DNA grafting density $\chi$.}
\end{figure*}

To develop building blocks with tuneable time-dependent binding strength and specificity, we started with two sets of particles coated with non-complementary single-stranded DNA, so that they initially do not assemble. We then used the Primer Exchange Reaction to append onto one set of particles---which we call precursors---new DNA domains that are complementary to the DNA on the other set of particles---which we call co-assemblers (Fig.~\ref{Fig1}b,c). The strength of DNA-mediated interactions between two particles increases with the number of DNA strands with complementary sequences on each particle, $\chi$~\cite{Uberti2016,Cui2022,leunissen2010,Dreyfus2009}, so the particles will start to aggregate once a threshold conversion, $\chi_T$, is reached. The primer exchange reaction thus converts precursors into assembly-active building blocks. With this design we hypothesized that, at constant temperature, the particles cannot bind for a duration determined by the rate of the DNA polymerization reaction, after which they assemble via selective, DNA-mediated interactions (Fig.~\ref{Fig2}a).

To test this hypothesis, we mixed particles coated with the 10-nucleotide sequence $a$ with particles coated with the 9-nucleotide sequence $b$, DNA polymerase, nucleotides, and $0.1~\nM$ of the PER template $T_{a\rightarrow b'}$ that catalyzes the growth of 9-nucleotide domain $b'$ onto DNA with domain $a$ (see Methods for details). Domain $b'$, which we call the binding domain, is complementary to sequence $b$ and facilitates binding between the precursors and co-assemblers (see SI section 1 for the sequences). We found that, for approximately 40 minutes after mixing, the colloids did not bind. Then, the first particles began to aggregate and after 70 minutes nearly all particles had aggregated into fractal-like clusters (Fig.~\ref{Fig2}a). By contrast, particles coated with DNA sequences that are complementary from the start aggregate within seconds to minutes after mixing, limited only by the time it takes for particles to find each other~\cite{leunissen2010}. The results in (Fig.~\ref{Fig2}a) thus show that the DNA polymerization reaction on precursors result in the delayed onset of aggregation.

The results in Fig.~\ref{Fig2} suggest that the assembly time was limited by the time it takes the polymerization reaction to convert precursors to assembly-active building blocks, which is in turn controlled (for high enzyme concentrations) by the template concentration. We thus attempted to tune the delay to aggregation (\textit{i.e.} when approximately half the particles are aggregated) by varying the template concentration between $10~\nM$ and $20~\pM$. To measure the time at which aggregation began for different template concentrations, we recorded movies of the assembly of particles in the presence of these different template concentrations. To quantify the degree of aggregation as a function of time, we binarized and inverted the microscopy images and measured the number of clusters, or particle aggregates, which in the processed images consisted of connected areas of white pixels. Figure~\ref{Fig2}b shows the number of clusters normalized to the number of clusters in the first frame plotted as a function of time. In each case, the number of clusters remained constant for a duration that depended on the template concentration and then decreased over time, indicating a delayed aggregation process. The average cluster size as a function of time showed a similar trend (Supplementary Figure 1).

To quantify the delay after which aggregation occurred, we measured the time until the number of clusters decreased to half the number of clusters in the initial frame; we call this the lag time. The lag time was less than 5 minutes for a template concentration of $10~\nM$, and over 8 hours for a template concentration of $20~\pM$. Figure~\ref{Fig2}c shows that the lag time was roughly inversely proportional to the template concentration, indicating that it can be systematically varied over a wide range by controlling the template concentration.

We further observed that the lag time can be predicted from the DNA polymerization rate and a threshold fraction of binding domains on the particle, $\chi_T$, that leads to binding. In our previous work~\cite{Moerman2023} we measured that the conversion of the DNA strands on the particle follows an exponential curve with a typical reaction time $\tau$ that is roughly inversely proportional to the template concentration, $c_T$. Given that $\delta t$ is the time it takes for the number of binding domains to reach the threshold required for assembly, and assuming that this number is small compared to the total number of DNA strands on the precursor particle, the lag time scales with the template concentration as (see Supplementary Note 2 for the derivation): 
\begin{equation}
\label{lagtimeprediction}
\delta t=-\frac{\tau_0 \ln{\left ( 1-\chi_T \right )}}{c_T}
\end{equation}
Here, the time scale $\tau_0$ is the inherent polymerization rate with the template concentration, $c_T$, scaled out so that $\tau=\tau_0/c_T$. $\tau_0=1.2 \times 10^3 \nM \min$~\cite{Moerman2023} for the sequences used in this paper. Fitting Equation~\ref{lagtimeprediction} to the data in Figure~\ref{Fig2} gives a value for the threshold  number of DNA strands required for aggregation: $\chi_T \approx 7\times 10^{-3}$. That is, assembly should start when only 0.7\% of the DNA strands per particle are extended so that they present a binding domain. The particles present approximately $10^5$ DNA strands per $\um^2$, so on the order of $10^3$ strands per 1-\um~particle are required for assembly under our experimental conditions, consistent with previous measurements~\cite{Dreyfus2009,Cui2022}.

Taken together, this analysis implies that DNA-coated particles with non-complementary sequences could function as precursors of self-assembly that are converted into assembly-active building blocks after a specific time delay using a template-catalyzed DNA polymerization reaction. The template concentration controls the lag time, which can range from minutes to multiple hours and the relationship between the time delay and the template concentration can be quantitatively understood from simple reaction kinetics---it can be predicted when the polymerization rate $\tau_0$ and the threshold binder density $\chi_T$ are known.

\subsection*{Multi-stage assembly}
\begin{figure*}
\includegraphics[scale=1]{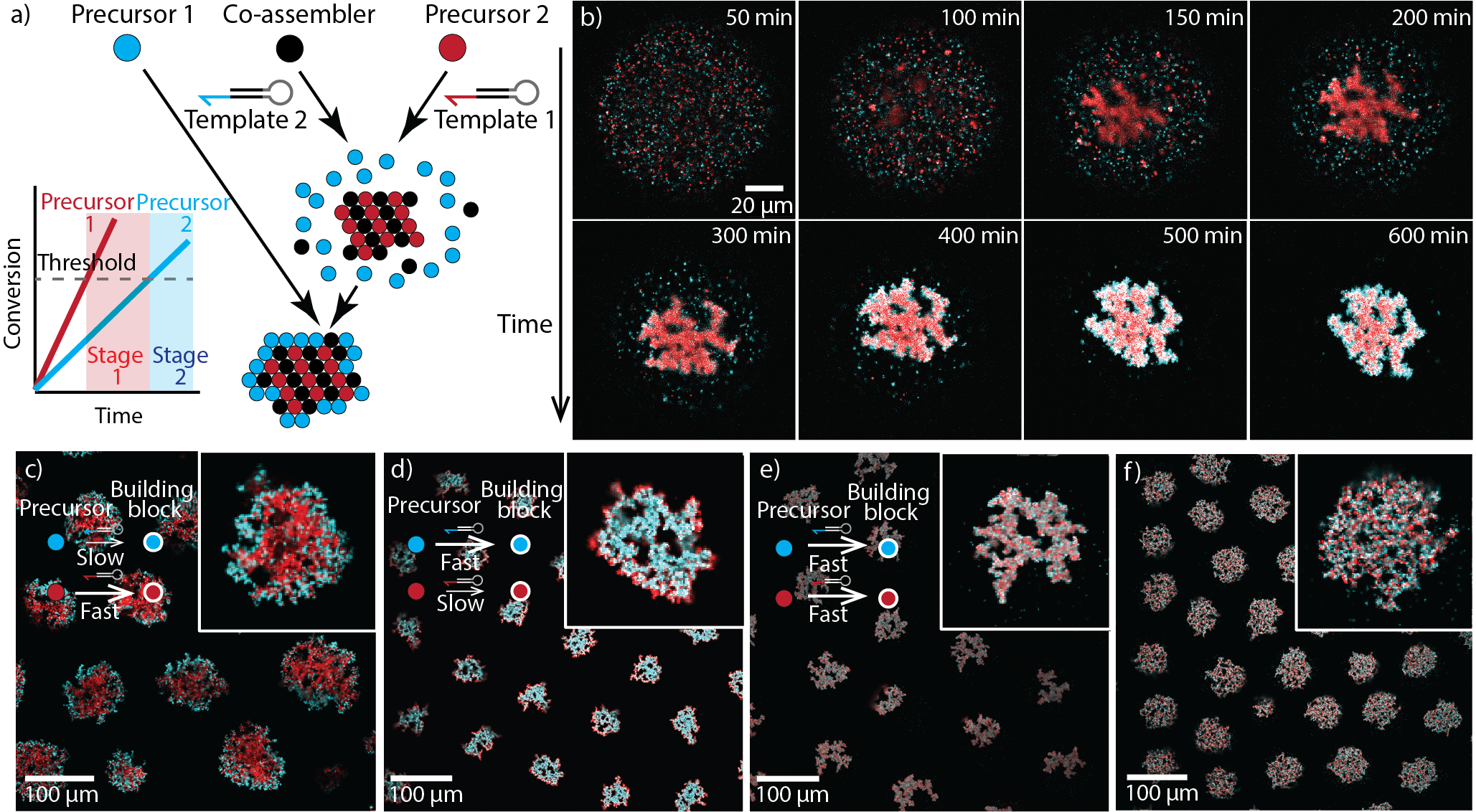}
\caption{\label{Fig4} a) The spatial heterogeneity of an assembly can be encoded in the assembly kinetics by controlling when particles are activated for assembly. Here, red particles are quickly activated---and thus able to aggregate with the co-assembler---via PER because template 1's concentration is relatively high, whereas the blue particles are activated after a longer delay because template 2's concentration is low. The graph (bottom left) sketches how the relative speeds of conversion of the DNA strands on the precursor particles leads to the emergence of two distinct assembly stages. b) Time evolution of cluster assembly where the reaction converting the red particles into assembly-active building blocks is fast (template concentration~$=1~\nM$) and the reaction converting the blue particles into assembly-active building blocks is slow (template concentration~$=0.1~\nM$). The consequence of the disparate reaction rates is that the red particles start to aggregate after roughly 1 hour, while the blue particles only start to aggregate after roughly 5 hours. c) Assemblies formed following the schematic in (a) have heterogeneous composition; the core contains more red particles while the shell contains more blue particles. d-e)  Clusters of the same particle composition as in (b,c) but following different reaction kinetics. The insets show zoom-ins of one of the clusters. d) When the blue particles aggregate first, and the red particles second, core-shell structures from with a blue core and a red shell. e) When both particles aggregate simultaneously, they form mixed clusters. f) Homogeneous clusters formed after core-shell clusters were heated above the temperature where the DNA-coated particles unbind ($60\C$) and subsequent cooling. After heating and cooling, the clusters do not regain the compositional heterogeneity that results from the two distinct assembly stages encoded in the DNA recipe.}
\end{figure*}
We next asked how the ability to tune the delay preceding particle assembly can be leveraged to orchestrate the assembly kinetics of multiple types of building blocks into kinetically trapped structures. To understand how tunable aggregation delays can determine the type of structure that is assembled, we studied how the co-assembly of three types of particles could result in final structures with different extents of compositional heterogeneity depending on their delays before assembling. The particles---precursor 1, precursor 2, and the co-assembler---carry non-complementary sequences $a$, $c$, and $b$ respectively. Two templates for PER, $T_{a\rightarrow b'}$ and $T_{c\rightarrow b'}$ append new domains, $b'$, onto the precursor particles that allow them to bind to the co-assembler. After the PER reaction completes, the now assembly-active particles 1 and 2 carry the same number of binding domains $b'$ so that they bind equally strongly to the precursor particles. Given that the number of DNA strands per particle, the sequence of the binding domain, and the number of particles are the same, the equilibrium state is a mixed crystal of all three particles with no spatial variation in the composition. However, depending on the relative rates with which the binder domains $b'$ are added to the precursors, the assembly of one of the precursors with the co-assembler can---during the reaction---transiently be faster than the assembly of the other precursor with the co-assembler. The assembly kinetics can thus be controlled by setting the concentrations of the templates $T_{a\rightarrow b'}$ and $T_{c\rightarrow b'}$ to create distinct assembly stages in which first one of the precursors is converted to an assembly-active particle, and so assembles, while the second precursor is converted and starts to assemble only after the first stage is complete (Fig.~\ref{Fig4}a). The precursor particles are fluorescently labeled to distinguish them and the co-assembler is unlabeled so that it is not visible in confocal fluorescence micrographs.

In an experiment where we mixed the two precursors, the co-assembler, $1~\nM$ of the PER template $T_{c\rightarrow b'}$, $0.1~\nM$ of the PER template $T_{a\rightarrow b'}$, DNA polymerase, and nucleotides, we observed a two-stage assembly process that resulted in clusters with heterogeneously distributed particle populations. The particles that assembled in the first stage were over-represented in the core, whereas the particles that assembled in the second stage were over-represented in the shell. To visualize---and later quantify---the heterogeneity of the assembled structures, we performed the assembly reactions in microfluidic water droplets in fluorinated oil~\cite{Hensley2023,hensley2022}, which resulted in finite-sized and roughly bowl-shaped assemblies (see Methods and Supplementary Figures 2,3). Figure~\ref{Fig4}b,c shows the structures that formed when the concentration of template that activated precursor 1 (blue) was $0.1~\nM$ and the concentration of template that activated the precursor 2 (red) was $1~\nM$. The structures have a predominantly red core and blue shell.

The separation in time of the two assembly stages, which resulted in the core-shell structure, is clearly visible in the time series in Fig.~\ref{Fig4}b. There was no aggregation for roughly the first 100 minutes of the experiment. Then, in a first stage of assembly, the red particles began to aggregate. The onset of aggregation of the blue particles was even later, starting after roughly 300 minutes.

Next we set out to show that the structure of the assemblies could be controlled by the assembly kinetics. To this end we prepared three samples with 1:1:1 ratios of the precursors 1 and 2 and the co-assemblers and added a different set of templates to all three samples (so that they follow a different molecular recipe for assembly): The first sample contained $1~\nM$ of the hairpin that converts the blue particles and $0.1~\nM$ of the hairpin that converts the red particles, and resulted in blue-core red-shell structures (Fig.~\ref{Fig4}d). The second sample contained $0.1~\nM$ of the hairpin that converts the blue particles and $1~\nM$ of the hairpin that converts the red particles, and resulted in red-core blue-shell structures (Fig.~\ref{Fig4}c). And the third sample contained $1~\nM$ of both templates and resulted in mixed aggregates without spatial heterogeneity (Fig.~\ref{Fig4}e). The same set of particles thus assembled into qualitatively different structures depending on the DNA instructions, even though the binding properties of all particles were identical in the three samples both at the beginning and at the end of the assembly process. The only difference was in the timing of the reactions. That is, the structure of the core-shell assemblies depends on the assembly pathway, not on the initial or final properties of the particles.

To confirm that the core-shell structure in Figure~\ref{Fig4}c corresponds to a meta-stable state and is not the thermodyamically favored state for the sample composition, we heated the sample to above the particles' dissociation temperature ($60\C$) and cooled it back down to room temperature. After the heating and cooling cycle the particles formed mixed aggregates with no heterogeneity in their composition (Fig.~\ref{Fig4}f). The mixed aggregates are still kinetically arrested but now lack the compositional heterogeneity that resulted from the DNA-encoded assembly kinetics. Notably, the structures in Fig.~\ref{Fig4}e and f differ in microstructure and compactness even though the densities of the red and the blue particles are uniform in both, indicating that the absolute reaction and assembly rates matter as well as the relative rates and are an important design consideration.

\subsection*{Delay determines layer composition}
\begin{figure*}
\includegraphics[scale=1]{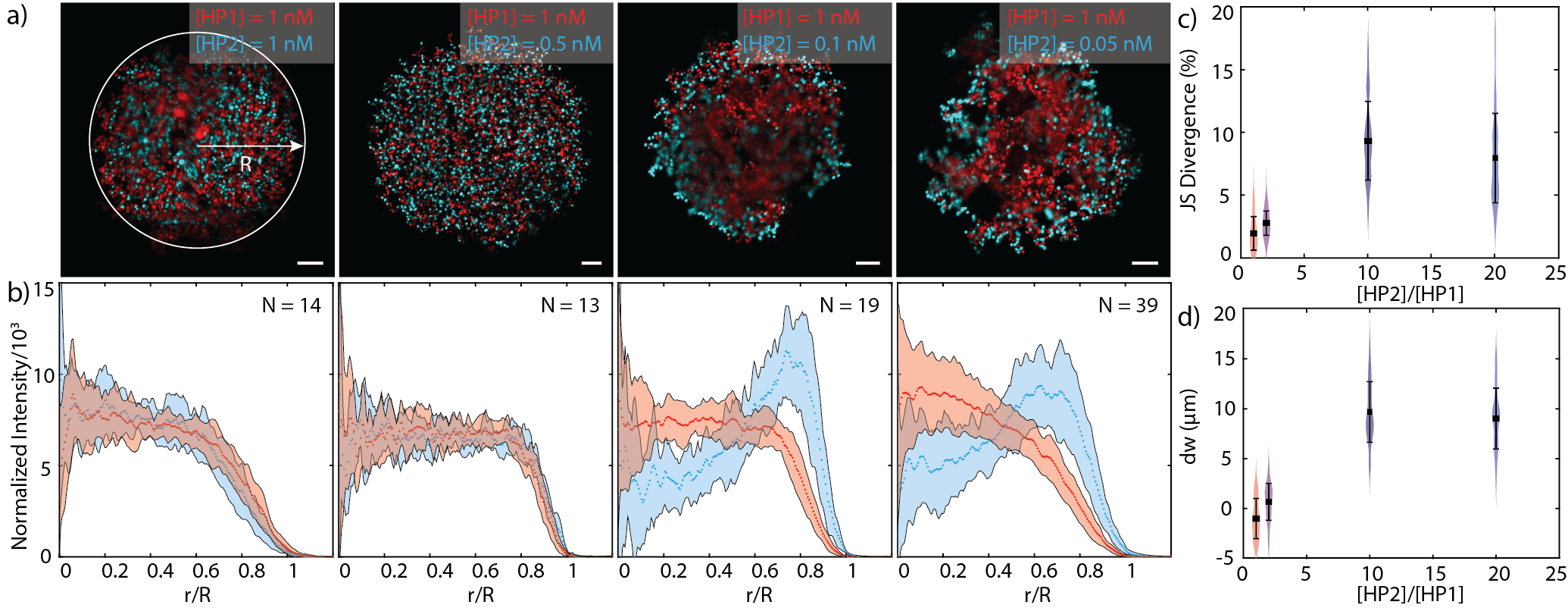}
\caption{\label{Fig6} a) Multi-component clusters self-assembled following different kinetic recipes. When the reactions that convert the two precursor particles have similar rates, the structures have no compositional heterogeneity. When one reaction is much faster than the other, the clusters' composition radially changes over length scales of approximately 10 times the building block diameter. The scale bars are $10~\um$. b) Radial composition of the cluster. The composition quantified using the fluorescent intensities of the red and blue particles as a function of distance from the cluster center. The y-axis is the radially averaged fluorescence intensity, normalized to the total intensity such that the area under the curve is 1. The x-axis is the radius of the cluster scaled to correct for small differences in cluster size (see Methods for details). The intensity profiles are averaged over $N$ clusters as given in the plots and the shaded area indicates the standard deviation around the average at each value of $r/R$. c) The Jensen-Shannon divergence, a measure of similarity, of the two profiles as a function of the ratio of the two template concentrations as a function of the concentration of the two template concentrations. The square marker and error bar indicate the average and standard deviation. The colored area indicates the distribution of the $N$ individual measurements. d) The distance, $dw$, from the edge of the red cluster to the edge of the blue cluster as a function of the ratio of the two template concentrations. $dw$ is a measure of the length scale over which the two particle populations are heterogeneously distributed across the cluster. The first two compositions result in homogeneous clusters with no significant difference between the two populations. The final two compositions have heterogeneities on order of $10~\um$, ten times larger than building block size. In both c and d the error bar represents the standard deviation and the violin plot shows the distribution of the measurements.}
\end{figure*}

To systematically study how the assembly kinetics affect the cluster composition, we varied the ratio of the template concentrations (Fig.~\ref{Fig6}a) and quantified the compositional heterogeneity of the resulting structures using the radial distributions of fluorescent intensity (Fig.~\ref{Fig6}b). We used the Jensen-Shannon (JS) divergence as a measure of spatial heterogeneity. This quantity is a measure of the difference between two distributions, normalized to the average distribution: $JS(P||Q)=H(M)-(H(P)+H(Q))/2$, where $M$ is the average of the two distributions $P$ and $Q$, and $H$ is the Shannon entropy~\cite{Lin1991}. In other words, it is the amount of information (\textit{i.e.} the Shannon entropy in bits) required to transform one distribution into another distribution, normalized to the information of the average distribution. It is thus related to the amount of order in a system~\cite{Martiniani2019}. We expect the binding energies of the co-assembler and the precursors 1 and 2 to be equal at the end of the reaction (when the activation of both particle types is completed), so that only the positioning of the particles contributes to the free energy differences between samples. The JS divergence is thus a qualitative measure of the free energy difference between the meta-stable state in which the system is trapped and a deeper minimum in the free energy landscape that corresponds to a mixed aggregate.

We found that the JS divergence was small (indicative of similar distributions of the two types of particles across the cluster and thus a low free energy) when the template concentrations---and by extension the reaction kinetics---are similar and that the JS divergence was larger (indicative of dissimilar distributions of the two types of particles across the cluster and thus a high free energy) when one template concentration is larger than the other (Fig.~\ref{Fig6}c). 

To quantify the length scale of the structures' compositional heterogeneity, we measured the difference between the typical size of the red part of the cluster and the typical size of the blue part of the cluster. We used the median of the red and blue distributions in Fig.~\ref{Fig6}b as a measure of the sizes of the respective red and blue parts of the cluster, and defined the distance between them, $dw$, as the position of the median of the blue distribution minus the position of the median of the red distribution. Fig.~\ref{Fig6}d shows that this distance follows a similar trend as a function of hairpin ratios to the JS divergence. 

The increases in both JS divergence and the red/blue cluster distance indicate that core-shell structures form when the delay between the two assembly stages is large enough and that mixed structures form when the delay is short. The composition of the structures also depends on other parameters than the aggregation delay, however, such as the initial particle composition (Supplementary Fig. 4). The spread in heterogeneity is large for a single condition (Fig.~\ref{Fig6}c,d), and even within a single experiment (Fig.~\ref{Fig4}d). 

\subsection*{Simulations}
\begin{figure}
\includegraphics[scale=1]{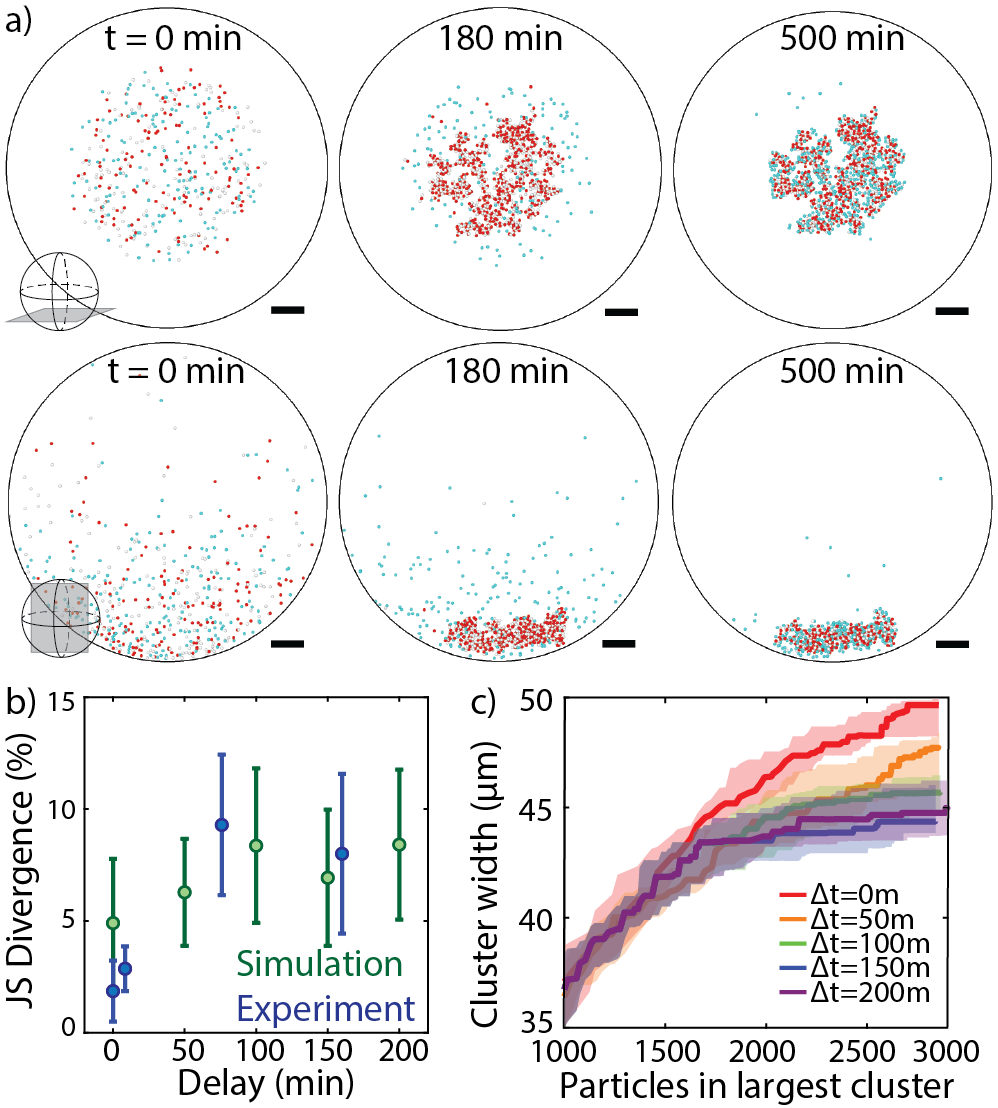}
\caption{\label{Sims} a) Snapshots of a simulation with a time delay of 180 minutes between the two assembly stages. The snapshots show that the red and white particles co-assemble in the first stage and the blue particles attach to the outside of the aggregates in the second stage in a sample simulation. The top row shows a $10~\um$ thick horizontal slice and the bottom row shows a $10~\um$ thick vertical slice. The snapshots of the particle aggregation, were rendered from the top and side using the 3D visualization tool, Ovito. The time delay between the assembly stages was 180 minutes. Scale bars are $10~\um$. b) The Jensen-Shannon divergence as a function of the time delay between the two assembly stages. The green markers indicate the Jensen-Shannon divergence from the simulations, calculated from the intensity of red and blue particles in the simulation. The error bars represent 1 standard deviation derived from 1000 bootstrap iterations, each sampling 25 points with replacement. The blue markers indicate the JS divergence from the experiments, as shown in Fig.~\ref{Fig6}c, with the x-axis converted to a time delay using~\eqref{lagtimeprediction}. c) Width of the largest cluster in the simulation as a function of the number of particles in the largest cluster for varying delay times. The width grows to about 50 particle diameters ($50~\um$) following a similar trend for all time delays. Simulations with shorter time delays resulted in slightly wider clusters.}
\end{figure}
To better understand how assembly delay can induce the formation of heterogeneous structures, we developed a simulation of time-delayed assembly using the Non-Equilibrium Reaction-Diffusion Simulations implemented in NERDSS (see Methods). To mimic the experiments, we simulated the assembly of 3000 particles in a 1:1:1 ratio between precursor 1 (red), precursor 2 (blue), and the co-assembler (gray), in a spherical simulation chamber with a diameter of $100~\um$. For simplicity, we assume that all particles of the same type can bind to others with a single attachment rate $k_{on}$ at a time prescribed by particle time and delay, and that binding is irreversible.

Snapshots of simulations with a delay of 200 minutes between the assembly stages (Fig.~\ref{Sims}a) show that the red particles, which aggregate in the first stage, form the core of an aggregate in the center of the sample chamber and that the blue particles, which aggregate in the second stage, stick to the outside of this cluster, consistent with experimental observations. The simulations also showed that the clusters are bowl-shaped with a height of about $15~\um$, following the contour of the bottom of the sample well, and that the blue particles attach evenly to the side, bottom, and top of the red aggregate and fill the interstitial spaces. The degree of heterogeneity quantified from a 2D projection of this structure is thus likely an underestimate of the compositional heterogeneity in 3D.

To investigate the effect of the delay between assembly stages on the final structure, we varied the delay between 0 minutes and 180 minutes -- the estimated longest delay in the experimental set -- and measured the JS divergence of the assemblies as a measure of their compositional heterogeneity. We found that the JS divergence increased with delay, consistent with our experimental findings (Fig.~\ref{Sims}b), though less so in the simulations than in the experiments. This difference may come from the fact that in simulations, every particle can be visualized, while in experiments particles deeper into the structure appear more blurry due to scattering effects.

We used the results of simulations, which allow for the direct observation of the assembly pathway and an assembly's 3D structure, to ask how assembly delays lead to compositional heterogeneity. As expected, in simulation, the assembly of red particles precedes that of blue (Fig.~\ref{Sims}a). In the simulations, clusters grew by the attachment of particles to the edges, the top, and the bottom of the cluster. The compositional heterogeneity in the horizontal plane arises due to the continued growth of the clusters outward, which creates blue exclusive domains. The attachment of blue particles more to the top and bottom of the clusters, later in assembly, also suggests how time delay could induce heterogeneity in the z-direction.

To compare the structural features of clusters that form with different, time-dependent rates, we used the number of particles in the largest cluster as a reaction coordinate (Supplementary Fig. 5). Surprisingly, we found that the structural features of the assembly, such as the cluster's width (Fig.~\ref{Sims}c), height (Supplementary Fig. 6), and fractal dimension (Supplementary Fig. 7), subtly depended on the time delay, $\Delta t$, between the assembly stages. This dependency suggests that varying the time delay between successive stages affects not only the compositional heterogeneity but also the overall structure of the clusters. One reason that time-delay could alter the cluster's structure is that it changes the effective concentrations of particles participating in the assembly reaction over time. Because the structures form by irreversible aggregation, reducing the concentration of particles able to assemble at the beginning of the reaction, for example, would make the structures more compact, \textit{i.e.} lowering their fractal dimension~\cite{gonzalez1999}. This prediction of the simulation is consistent with the observation that the structure of aggregates that formed after a heating and cooling cycle had a more open structure than aggregates that formed by slow conversion of their DNA coatings~\ref{Fig4}e. These results suggest how time-delay reactions might be used to control overall structure and assembly shape as well as compositional heterogeneity.

\section*{Discussion and conclusions}
In this article we showed that sets of DNA templates for an isothermal polymerization reaction can be used to selectively convert assembly precursors into assembly-active building blocks after prescribed time delays. This set of templates amounts to a DNA-encoded recipe that describes the assembly pathways to specific kinetically trapped structures. The pathway-dependent assembly into kinetically trapped structures is a type of non-equilibrium assembly, where the energy that drives the process is provided by the exergonic conversion of dNTPs to DNA oligomers.

Templates that control the conversion of precursors into assembly-active building blocks could conceivably be produced or degraded by upstream reaction networks~\cite{Fujii2013,Schaffter2019,Qian2011}, or diffuse from localized sources~\cite{Pistikou2023}, so that local concentrations of template locally change the binding properties of the building blocks and drive their assembly into non-equilibrium structures through controlled kinetics. This opens the way to orchestrating the assembly of a limited set of building blocks into a diverse set of structures by using local variations in the concentration of instruction DNA, analogous to cell differentiation in response to local morphogen concentrations. 

A fundamental difference between methods for transiently changing the binding properties of particles based on linker-mediated interactions~\cite{Zadorin2017,Dehne2021,Lowensohn2019} and our covalent approach is that the covalent changes to the particles’ DNA composition remain after the signal (\textit{i.e.}  template) is gone, providing the particles with a memory of their local chemical environment (\textit{i.e.}  the local template concentrations they experienced), while the particles’ responses to linker concentrations track fluctuations in the linker concentrations without delay. Linker-based approaches are advantageous when targeting transient and dynamic patterns that obey kinetics imposed by  \textit{e.g.} reaction networks, while covalent DNA polymerization is an advantage when driving the system into a final kinetically trapped state. The long term memory provided by covalent changes to the particles’  DNA coatings also open the door to engineering ``metaproperties'' such as adaptability and physical learning in colloidal materials~\cite{Stern2023,Falk2023}.

Our kinetic control knobs over the assembly pathways complement
and can be used in conjunction with geometric control knobs to direct self-assembly into target structures, such as particle shape and patchiness. One could imagine using DNA-encoded recipes to prescribe when different interaction patches are activated to design complex pathways to high-energy structures that are otherwise inaccessible by equilibrium self-assembly.

The DNA-encoded recipes that offer kinetic control over multi-step processes are useful not only in colloidal self-organization but also potentially for directing the folding of DNA origami~\cite{dunn2015} or colloidomers\cite{McMullen2022}, which notoriously happen in rugged energy landscapes with many local minima.

Our simulation tool allowed us to qualitatively investigate how control over time delays in assembly may be used to assemble structures with compositional heterogeneity over length scales much larger than the building block size. The ability to quantitatively predict experimental outcomes using simulations would make it possible to explore and then design the class of new structures enabled by time-delay assembly processes. Developing such predictive simulations for inverse design would require detailed modeling of the interaction potentials between particles, and how they change over time~\cite{Dijkstra2021,goodrich2021}; and a way to deal with reorganizations within clusters, which may be affected by hydrodynamics~\cite{tateno2019} and the multivalent nature of DNA-mediated interactions~\cite{hensley2022}.

Taken together our work is a step toward a framework that connects positional information encoded in the building block properties and their local short-ranged interaction potentials, and pathway information encoded in the assembly process, to give control over the final structure in both the microscale (on the order of the building block) and the mesoscale (larger than that of the building block but smaller than the system size).

\section*{Methods}
The oligonucleotides were purchased from Integrated DNA Technologies, in the "lab ready" formulation (dissolved in IDTE buffer at $100~\uM$). The unmodified DNA was ordered unpurified (\textit{i.e.} standard desalting). The DBCO-modified DNA (Integrated DNA Technologies) was HPLC-purified. The DNA sequences are listed in Supplementary Note 1.

All solutions were prepared in Millipore purified water. DNA was stored at $-20$ \degree{C} and DNA-coated particles were stored at $4$ \degree{C}

\subsection*{Synthesis of DNA-coated colloids}
We synthesized DNA-coated colloids as described in our previous work~\cite{Moerman2023} following the method described in detail in the ESI of ref.~\cite{Gehrels2022}, which is a modified version of the method originally put forward in ref.~\cite{Oh2015}. In brief, the method consists of three steps: 1) Azide groups are attached to the ends of polystyrene-poly(ethylene oxide) (PS-PEO) diblock copolymers; 2) The azide-modified PS-PEO copolymers are physically grafted to the surface of polystyrene colloids; and 3) Single-stranded DNA molecules are conjugated to the ends of the grafted PS-PEO copolymers by strain-promoted click chemistry. 

\subsection*{Primer exchange reaction}
To append new sequences onto the DNA grafted to the particles, we used an adapted version of the Primer Exchange Reaction (PER) described in Ref.\cite{Moerman2023}:
First we prepared a PER reaction mixture containing final concentrations of 1x Thermopol DNA polymerase buffer (New England Biolabs, provided at 10x concentration with the DNA polymerase), $12.5~\mM$ MgCl$_2$ (New England Biolabs, provided as a $100~\mM$ solution with the DNA polymerase), and $100~\uM$ of each dATP (deoxyribose adenine triphosphate, ThermoFisher, $10~\mM$), dCTP, and dTTP. No dGTP was added to the mixture because only ACT sequences were appended onto particles and a G-C pair was used as a stop sequence.

For the polymerization-directed self-assembly experiements we prepared samples of a total volume of $5~\uL$ containing $0.1$~wt\% of each of the DNA-coated particles participating in the assembly process; DNA template concentrations that varied between $0.01~\nM$ $and 1~\nM$ and determined the rate of the reaction; and $0.13~$U/$\uL$ Bst Large Fragment DNA polymerase (New England Biolabs). 

\subsection*{Microfluidic droplet generation}
For the experiments in Fig.~\ref{Fig4} and Fig.~\ref{Fig6}, where the degree of core-shellness in response to varying assembly kinetics was quantified, the clusters were assembled in water-in-oil emulsion droplet micro-chambers in order to distinguish a core and a shell. In larger chambers the particles formed space-spanning fractal networks for which it was harder to visualize and how the composition varied with distance to the edge of the cluster. To make monodisperse water-in-oil emulsion droplets, we followed the procedure described in ref.~\cite{Hensley2023,hensley2022}. In short, PDMS microfluidic channels with channel widths of $100~\um$ and channel heights of approximately $50~\um$ were produced. Two syringe pumps (Harvard instruments) pushed the aqueous mixture of the DNA-coated particles, buffer, DNA templates, nucleotides, and polymerase through the central channel with a flow rate of $300~\ulh$; and a 5 wt$\%$ fluorosurfactant (Fluorinert) solution in fluorinated oil (FC-40) through the outer channel with a flow rate of $500~\ulh$. The particle mixture had a total sample volume of $5~\uL$, which is too small to load into the syringes, so the sample was loaded into the the tube connecting the syringe to the microfluidic device with an air bubble separating it from water on either side. 

For experiments where the assembly needed to be imaged shortly after starting of the reaction, (SI Movies 1 and 2, Fig.~\ref{Fig4}b), the water-in-oil emulsion droplets as microchambers were prepared by combining the aqueous solution containing the DNA-coated particles, enzymes, nucleotides, buffer, and DNA templates with the fluorosurfactant in fluorinated oil solution in an Eppendorf tube and mixing by flicking the tube three times. The resulting size distribution of the droplets appeared suitable for imaging. 

\subsection*{Microscopy}
To prepare the microscope samples, droplets were collected in a 0.2 mm x 2 mm x 20 mm glass capillary (Invitrogen) and the capillary was glued closed and attached to a microscope slide with 2-component glue (Bison). Then we imaged the samples using either an inverted, brightfield microscope (Nikon Eclipse TE2000-E) (Fig.~\ref{Fig2}) or a confocal microscope (Leica SP8) (Fig.~\ref{Fig4}, Fig.~\ref{Fig6}) equipped with 20x air, 40x air, and a 60x oil immersion objectives.

\subsection*{Simulation}

All simulations were performed using the Non-Equilibrium Reaction-Diffusion Self-Assembly Simulator (NERDSS) as described in ref.~\cite{varga2020}. NERDSS is a computational tool designed to simulate the self-assembly of biological and synthetic systems at the mesoscale. It integrates spatially resolved reaction-diffusion models with structural resolution, enabling the capture of particle diffusion, binding, and unbinding dynamics over long timescales and large system sizes.

To model DNA-coated colloid assembly, the NERDSS\cite{varga2020} framework was modified to incorporate mechanisms specific to DNA-coated colloid interactions. DNA-coated colloidal particles were represented with binding sites to simulate specific hybridization events. The translational and rotational diffusivities of the particles were calculated using the Stokes-Einstein law. Primer Exchange Reaction (PER) was modeled as a time-dependent activation process with a fixed binding constant $k_a=1 nm^3 \mu s^{-1}$, allowing for programmed assembly at various stages. NERDSS was adapted to operate on micrometer (µm) and millisecond (ms) scales, suitable for colloidal systems that span hours, rather than the nanometer (nm) and microsecond (µs) scales typically used in biological systems. Binding rules were implemented to enumerate binding events between individual sites, accurately representing the dynamics of "hit-and-stick" hybridization between colloids. Spatial constraints were enhanced to account for the excluded volume and rigid structures of DNA-coated colloids. Additionally, settling terms were introduced to simulate sedimentation equilibrium, enabling the modeling of aggregation as particles settle and interact in a gravitational field. The simulation outputs were processed to produce emulated confocal micrographs for comparison with experimental data by selecting a z-segment of the images and blurring particles to reflect focus effects. For full simulator description, please refer to our modeling study ref.~\cite{Chou2025prep}. Code is available at https://github.com/s106030605/NERDSS\_on\_DNA \_colloid/tree/main.

\subsection*{Data analysis}
All analysis was done with Matlab version R2021a.

To measure the average cluster size as a function of time shown in Figure~\ref{Fig2}b, we first binarized and inverted the images such that pixels containing a particle had value 1 (white) and pixels containing no particle had a value 0 (black). Then we used Matlab's `bwconncomp' function to find all clusters of connected white pixels and measure their sizes. Figure~\ref{Fig2}b shows the average number of pixels per connected normalized to that value in the first frame.

To measure the normalized intensity distributions in Figure~\ref{Fig6}b, we first made a composite image containing the average intensity and the red and blue channel. We identified the center of each of the clusters in the composite image by blurring it and using Matlab's `bwconncomp' function. We then removed objects that were misidentified as individual clusters, that were out-of-focus, or that overlapped with the image edge. Next, for each cluster, we radially averaged the red and blue pixel intensity separately, we identified the edge of the cluster as the distance from the center where both the red and blue pixel intensity had dropped below an arbitrary threshold value of 100 counts (we empirically found this was the appropriate cut-off value), and normalized the intensity so that the area under the curve equals 1. The graphs in Fig~\ref{Fig6}b show the average of all graphs of clusters in a single condition.

We calculated the JS divergence shown in Figure~\ref{Fig6}c from the normalized and scaled intensity profiles in Figure~\ref{Fig6}b, $P(r)$ and $Q(r)$ where P and Q are the red and blue intensities respectively. We first calculated the average distribution $M(r)$; for each $r$ $M(r)$ is the average of $P(r)$ and $Q(r)$. Then we measured Kullbeck-Leibler divergence for the two distributions with respect to the average distribution: $KL(P||M)=Q$log2$(Q/M)$. Then we calculated the Jensen-Shannon divergence as $JS(P||Q)=JS(Q||P)=1/2((KL(P||M)+KL(Q||M))$

We calculated the difference in the widths of the blue and red clusters, $dw$, plotted in Figure~\ref{Fig6}d by comparing the medians of the normalized red and blue distributions in Figure~\ref{Fig6}b. We measured $dw$ by subtracting the $r$ that indicates the median of the red distribution from the $r$ that indicates the median of the blue distribution and multiplying it by the size of the cluster, $R$. 

To demonstrate a statistically significant difference, we employed bootstrapping with 1,000 resampled datasets, each containing 25 replacement samples. Bootstrapping is a resampling technique that estimates the sampling distribution of a statistic by repeatedly drawing samples with replacement from the observed data. For hypothesis testing, we evaluated the overlap within a confidence interval defined by 1 standard deviation. If the confidence intervals of the compared groups did not overlap, it indicated a statistically significant difference, providing robust evidence to support our findings.

\section*{Acknowledgements}
P.G.M. and R.S. acknowledge funding from the ARO (grant number W911NF2010057) and the DOE (grant number DE-SC0010426), which supported the analysis of kinetic trajectories and assessment of structural heterogeneity in both experiments and simulation. C.C. acknowledges funding from the Department of Int. and Cross-strait Education. P.G.M. acknowledges funding by the American Institute of Physics through the Robert H.G. Helleman Memorial Fellowship. T.E.V., and W.B.R. acknowledge NSF funding (grant number DMR-1710112) and funding from the Brandeis Bioinspired Soft Materials MRSEC (grant number DMR-2011846). W.B.R acknowledges support from the Smith Family Foundation.

\section*{Author contributions}
P.G.M and R.S. conceived the experiment. P.G.M and T.E.V. performed the experiments and analyzed the data. C.C. performed the simulations and analyzed the data. All authors discussed results and contributed to writing of the manuscript.

\bibliography{mainrefs}

\end {document}